\documentclass[sigconf]{acmart}
\usepackage{graphicx} 

\title{OrchVis: Hierarchical Multi-Agent Orchestration for Human Oversight}
\author{Jieyu Zhou}
\affiliation{%
  \institution{Georgia Institute of Technology}
  \city{Atlanta}
  \state{Georgia}
  \country{USA}
}
\email{jzhou625@gatech.edu}
\date{October 2025}

\begin{document}
\begin{teaserfigure}
  \centering
  \includegraphics[width=\textwidth]{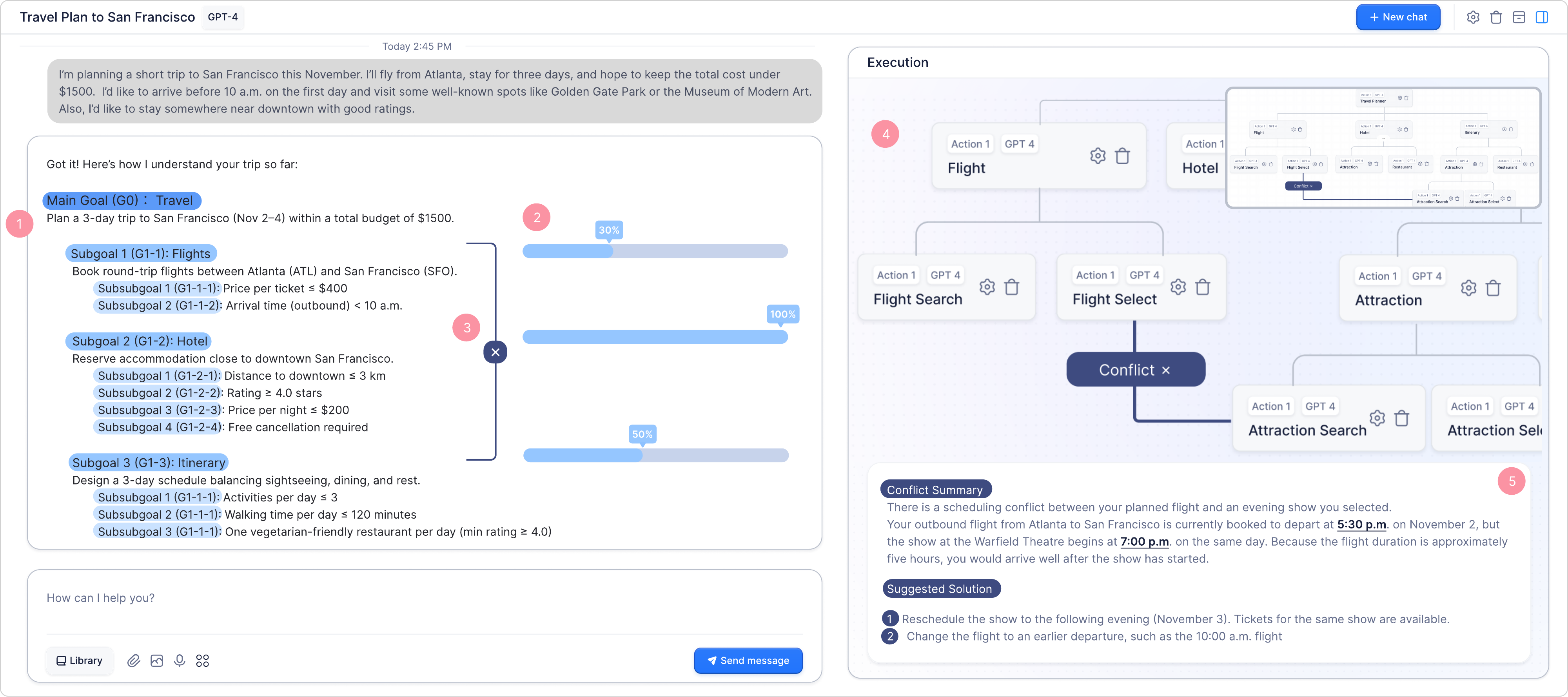}
  \caption{System overview of our multi-agent, goal-aware visualization. (1) In the planning phase, the user specifies the task; the orchestration/supervisor agent interprets the intent and decomposes it into structured subgoals. (2) During execution, the UI tracks and displays per-goal progress. (3) When inter-goal conflicts arise, they are flagged in situ and can be expanded into the Planning Panel. (4) The Planning Panel reveals how the orchestrator routes work to sub-agents and pinpoints which sub-agents are in conflict. (5) A summary pane presents high-level rationales, proposed resolutions, and predicted future states to help the user make informed decisions and set the desired level of autonomy.}
  \label{fig:pro}
\end{teaserfigure}

\begin{abstract}
We introduce \emph{OrchVis}, a multi-agent orchestration framework that visualizes, verifies, and coordinates goal-driven collaboration among LLM-based agents. Through hierarchical goal alignment, task assignment, and conflict resolution, OrchVis enables humans to supervise complex multi-agent workflows without micromanaging each step. The system parses user intent into structured goals, monitors execution via automated verification, and exposes inter-agent dependencies through an interactive planning panel. When conflicts arise, users can explore system-proposed alternatives and selectively replan. OrchVis advances human-centered design for multi-agent systems by combining transparent visualization with adaptive autonomy.
\end{abstract}
\maketitle

\section{Introduction}
Most commercial AI agents today execute tasks end-to-end with little to no user oversight \cite{openai2025agent, anthropic2024claude, manus2024}. In single-agent settings, HCI work has introduced a broad toolkit for human involvement in planning and execution \cite{shao2024collaborative, das2024vime, zhao2024lightva, suh2024luminate, feng2024cocoa, lawley2023val, epperson2025interactive, kazemitabaar2024improving}, such as co-planning and co-execution of research tasks \cite{feng2024cocoa}, pausing and editing agent behaviors in programming \cite{epperson2025interactive}, and decomposing complex data analysis workflows for verification By contrast, user oversight in multi-agent scenarios remains underexplored: most prior work emphasizes technical orchestration frameworks \cite{wu2024autogen, fourney2024magentic} or quantifies how human involvement scales with agent collectives \cite{7299280, clarke2024one}, with comparatively few system-level designs that make coordination legible and controllable to end users. One notable exception is CommunityBot \cite{jiang2023communitybots}, which focuses on socially oriented chat; however, it does not address the challenges of supervising multi-agent collaboration on complex task execution.

In this paper, we present OrchVis, a visualization and orchestration framework that enables users to oversee, understand, and correct multi-agent collaborations through hierarchical goal alignment, layered visualization, and conflict resolution. Building on insights from prior HCI work on mixed-initiative interaction, OrchVis rethinks how users can strategically guide multiple agents without micromanaging them. We first derive three design goals for enabling effective human oversight in multi-agent systems, then introduce the system design and technical implementation that realize these goals.

\section{Design Goals}
Based on prior research, we conclude three design goals for the multi-agent system.

\subsection{From Direct Manipulation to Orchestration.} There has been a long-standing debate in HCI: should agents operate under direct human supervision and control, or should they function in a mostly autonomous fashion \cite{horvitz1999principles, shneiderman1997direct, lieberman1997autonomous, maes1995agents}? However, for complex tasks involving multiple agents, direct manipulation quickly becomes infeasible. \citet{7299280} introduced the concept of human cognitive complexity, showing that directly coordinating $k$ sub-group agents scales at least as $O(k)$ in user effort, which grows rapidly as the number of agents increases. In contrast, an orchestration agent that supervise sub agents can substantially reduce this complexity . User studies further confirm that users consistently prefer orchestration over direct manipulation of multiple agents \cite{clarke2024one}. Some research \cite{moran1997multimodal} has conceptualized the human as simply another agent within the system. However, we argue that the human should take the lead, acting at a higher level of abstraction, supervising with autonomy.

\textbf{Design Goal 1.} Introduce an orchestration agent that supervises sub-agents and consolidates low-level information into high-level summaries for human oversight. The human, in turn, should operate at a strategic level rather than managing procedural details.

\subsection{Separate Goals from Tasks and Align Early.} We distinguish two forms of planning organization following prior work: goals and tasks (e.g., hierarchical task networks \cite{nau2003shop2}, and hierarchical goal networks \cite{shivashankar2013hierarchical}). A goal is a desired target state—a set of constraints that should hold in the world (the what). A task is an executable procedure—a sequence of actions that can make those constraints true (the how). For example, in flight purchasing, “have a confirmed ATL→SFO ticket under \$400” is a goal; searching flights, applying coupon, and checkouting are subtasks that may achieve it. 

Misalignment between the user’s intended goal and the agent’s inferred goal is common \cite{bansal2024challenges,coscia2025ongoal}, caused by vague user inputs and model interpretation drift \cite{li2020multi}. Detecting such misalignment early prevents costly derailments during execution \cite{zhou2025should}. In practice, users focus on outcomes first (goal satisfaction) and only examine process details (tasks) when results appear incorrect \cite{zhou2025should}. To reach the first design goal of showing right amount of information to users, unnecessary information-such as action sequences when the goal has benn achieved-should remain hidden. 

\textbf{Design Goal 2.} Before executing the task, ensure both users and agents agree on the goal. During execution, emphasize goal achievement as the primary view, while task-level details are revealed only when users need to inspect or correct errors.

\subsection{Visualizing Inter-Agent Relationship} The relationships between agents are often complex, essentially reflecting the intricate dependencies among subtasks within a larger task. Prior research has characterized task relationships as sequential, parallel, nested, or hierarchical \cite{wu2024autogen}. Agents may also engage in debate, which has been shown to improve overall task success rates \cite{du2023improving}. Meanwhile, prior work has found that text-heavy interfaces reduce users’ flexibility to revise or provide feedback, motivating the design of GUI-first interfaces for single-agent systems \cite{feng2024cocoa, arawjo2024chainforge}. For multi-agent systems, where dependencies and interactions among agents are even more intricate, GUI-first visualization becomes especially crucial for helping users understand and oversee agent relationships.

\textbf{Design Goal 3.} Visualize inter-agent dependencies and conflicts through a GUI to support user oversight.

\section{OrchVis: System and Technical Design}

\begin{figure}
  \includegraphics[width=\linewidth]{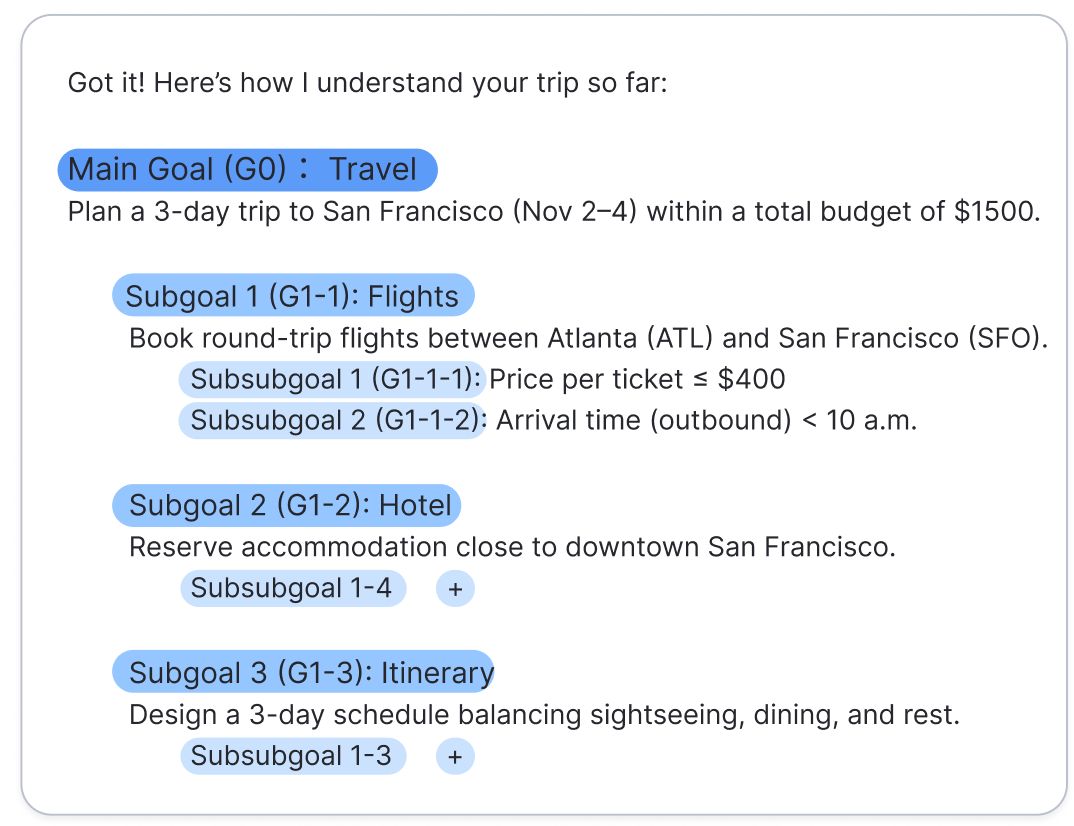}
  \caption{The user’s San Francisco travel goal parsed into flight, hotel, and itinerary subgoals. Lower levels can be expanded on demand.}
  \label{fig:goal_overview}
\end{figure}

OrchVis visualizes and coordinates multi-agent planning through a layered interface that balances automation with user oversight. 
Using a travel-planning scenario (Fig.~\ref{fig:goal_overview}), the system proceeds through four tightly coupled stages—goal alignment, task assignment, progress tracking, and conflict resolution—each supported by specific technical components (goal parser, orchestrator, verifier, and re-planner). 
Together, these stages enable users to monitor, diagnose, and revise multi-agent workflows while maintaining transparency and control across hierarchical goals and actions.

\subsection{Goal Alignment and Parsing}

The process begins when the user describes a task in natural language. 
The orchestration agent interprets this description and proposes an initial goal hierarchy. 
To reduce cognitive load, OrchVis displays only top-level goals (e.g., flight, hotel, itinerary), while lower-level goals remain collapsed but can be expanded for inspection. 
Users can directly edit goal texts, constraints, or subgoals before execution begins, ensuring alignment between user intent and the system’s interpretation.

Internally, the \emph{goal parser} converts free-form text into a structured goal graph using few-shot exemplars and grammar-constrained decoding. 
It identifies hierarchical relationships (sequential, parallel, or conditional), fills domain-specific fields, and normalizes attributes such as time and cost. 
Each goal node is grounded to an ontology and equipped with machine-checkable success predicates, enabling later verification. 
A lightweight repair loop ensures consistency between the user-edited hierarchy and the system’s internal representation.

\subsection{Task Assignment and Plan Generation}

After the goals are confirmed, the orchestration agent constructs a workflow that assigns subgoals to specialized sub-agents. 
This process is hidden by default to reduce cognitive load but can be revealed for inspection or manual adjustment. 
The orchestrator generates an executable plan and performs deterministic agent matching, comparing each task’s declared requirements against every sub-agent’s skill matrix that records its available tools, I/O schema, and historical performance. 
Sub-agents are LLM- or tool-augmented for specialized functions such as search, retrieval, or verification, and the resulting plan is represented as a composable task graph. 

\subsection{Progress Tracking and Goal Verification}

During execution, OrchVis continuously visualizes per-goal progress through the goal verifier. 
When conflicting or incomplete results arise—such as scheduling inconsistencies—the affected goals are flagged for user review. 
Users can expand the planning panel to inspect these goals in detail and view how sub-agents’ outputs map to goal completion.

The verifier evaluates execution states against the machine-checkable predicates generated earlier. 
For each subgoal, it checks hard and soft constraints and reports both a binary achievement flag and an overall satisfaction score:
\[
S = \frac{\#\text{satisfied hard}}{\#\text{total hard}} 
+ \lambda \times \frac{\#\text{satisfied soft}}{\#\text{total soft}}.
\]
Differences are summarized into structured reports that drive visualization updates and user notifications. 
When evidence is unstructured (e.g., textual or image data), a lightweight extractor LLM converts it into structured form for rule-based comparison.

\subsection{Conflict Detection and Planning Panel Correction}
\begin{figure}
  \includegraphics[width=\linewidth]{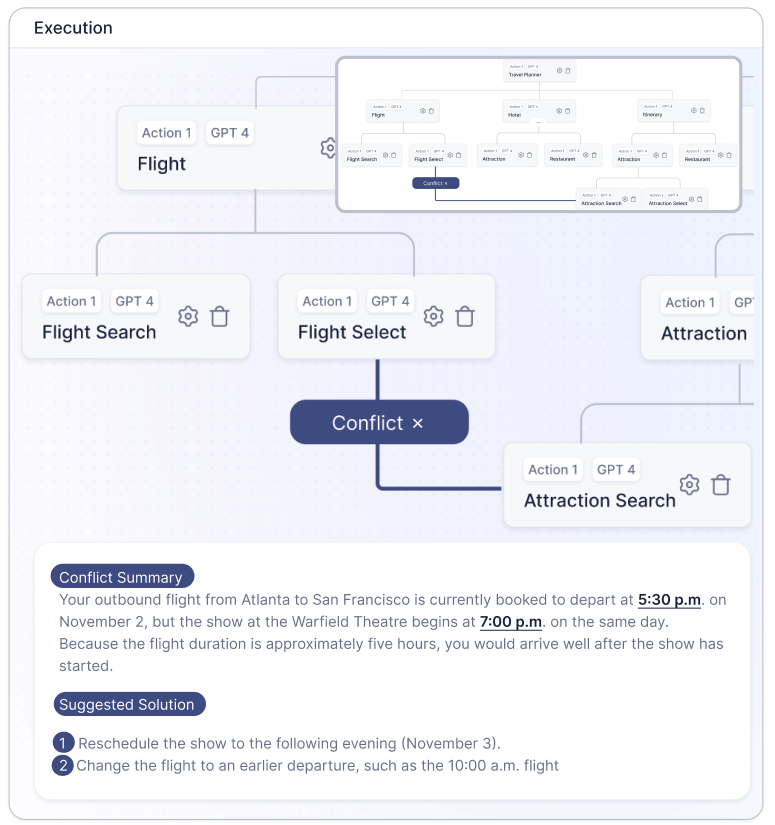}
  \caption{Detected conflict between the outbound flight and the evening show. Highlighted nodes indicate conflicting goals; the text panel below lists suggested repairs.}
  \label{fig:plan_conflict}
\end{figure}

When conflicts are detected, OrchVis expands the \emph{Planning Panel}, revealing a two-layer visualization. 
The upper layer shows the goal hierarchy, while the lower layer depicts the corresponding task-level workflows and tool calls. 
Users can explore alternative fixes, view predicted outcomes such as progress, risk, and cost, and select preferred solutions directly within the panel. 
The orchestrator pauses only affected branches while allowing other workflows to continue executing in parallel.

Upon receiving the verifier’s difference report, the conflict resolver translates it into natural language summaries and proposed repairs. 
If the user accepts a suggestion, the plan is updated and execution resumes; otherwise, the system triggers partial re-planning, generating alternative candidates via LLM-guided reasoning. 
All updates are propagated in real time to both the visual and textual layers to maintain transparency and continuity.

\section{Conclusion and Future Work}
This paper presents OrchVis, a visualization and orchestration framework for multi-agent systems that integrates goal alignment, task assignment, progress verification, and conflict resolution within a unified interface.

In future work, we plan to evaluate OrchVis across different benchmark environments to examine its adaptability to diverse task structures, such as online shopping, office assistance, and data analysis \cite{xie2024osworld, xu2024theagentcompanybenchmarkingllmagents, ma2024spreadsheetbench, jing2024dsbench, yao2022webshop, wei2025browsecomp}. We will also conduct user studies to assess how well the system supports human oversight and decision efficiency in long-horizon, multi-agent tasks.

\bibliographystyle{ACM-Reference-Format}
\bibliography{sources}
\end{document}